\documentclass[vecphys]{svmult}

\usepackage{makeidx}         % allows index generation
\usepackage{graphicx}        % standard LaTeX graphics tool
\usepackage{multicol}        % used for the two-column index
\usepackage[bottom]{footmisc}% places footnotes at page bottom
\usepackage{natbib}% places footnotes at page bottom

\def\apj{ApJ}           % Astrophysical Journal
\def\apjl{ApJ}          % Astrophysical Journal, Letters
\def\aap{A\&A}          % Astronomy and Astrophysics
\def\mnras{MNRAS}       % Monthly Notices of the RAS
\def\pasj{PASJ}         % Publications of the ASJ
\def\oiii{[O\,{\sc iii}]}
\def\hbeta{\hbox{H$\beta$}}
\def\nii{\hbox{[N\,{\sc ii}]}}

\def\kms{$\mbox{km s}^{-1}$}
\newcommand{\sauron}{{\texttt {SAURON}}}

\makeindex             % used for the subject index
\begin{document}

%\title*{\sauron\ observations of Sa spiral bulges:
%\newline the interacting galaxy NGC\,5953}

\title*{\sauron\ observations of Sa bulges: the formation of a kinematically
decoupled core in NGC\,5953}
\authorrunning{Falc\'on-Barroso et al.} 
\titlerunning{The formation of a KDC in NGC\,5953}

\author{Jes\'us Falc\'on-Barroso\inst{1}\and R. Bacon\inst{2}\and 
M. Bureau\inst{3}\and M. Cappellari\inst{1}\and \\
R.~L. Davies\inst{3}\and P.~T. de Zeeuw\inst{1}\and E. Emsellem\inst{2}\and 
K. Fathi\inst{4}\and D. Krajnovi\'{c}\inst{3}\and \\
H. Kuntschner\inst{5}\and R.~M. McDermid\inst{1}\and R.~F. Peletier\inst{6}\and 
M. Sarzi\inst{3,7}}

\institute{Sterrewacht Leiden, Leiden, The Netherlands.\newline 
\texttt{jfalcon@strw.leidenuniv.nl}
\and 
CRAL - Observatoire de Lyon, Lyon, France.
\and
University of Oxford, Oxford, United Kingdom.
\and
Rochester Institute of Technology, New York, USA.
\and
European Southern Observatory, Garching, Germany.
\and
Kapteyn Astronomical Institute, Groningen, The Netherlands.
\and
University of Hertfordshire. Hatfield. United Kingdom.}

\maketitle

%------------------------------------------------------------------
\begin{abstract}
We present results from our ongoing effort to understand the nature and
evolution of nearby galaxies using the \sauron\ integral-field spectrograph. In
this proceeding we focus on the study of the particular case formed by the
interacting galaxies NGC\,5953 and NGC\,5954. We present stellar and gas
kinematics of the central regions of NGC\,5953. We use a simple procedure to
determine the age of the stellar populations in the central regions and argue
that we may be witnessing the formation of a kinematically decoupled component
(hereafter KDC) from cold gas being acquired during the ongoing interaction with
NGC\,5954. 
\end{abstract}

%------------------------------------------------------------------
\section{Interacting galaxies and the formation of KDCs}
The interaction between galaxies lies at the heart of the current lore of
hierarchical formation models. While most of the recent numerical simulations
are able to make predictions about how such interactions affect the outer
regions of galaxies (e.g. Combes et al. 1995; Moore et al. 1999), it is
only in the last years that they have started to achieve sufficient resolution
to provide clues of their effects on the inner regions of galaxies (e.g. Naab \&
Burkert 2001). Signatures of counter-rotation in either the stellar or gas
components of galaxies have often been used to argue for the external origin of
material and therefore interaction between galaxies (e.g. Bertola et al. 1988).
In merger remnants, direct evidence for past interactions is the presence of a
KDC in the inner regions of the galaxy. 

One of the major goals of the \sauron\ survey is to study the structural and
kinematical properties of such KDCs, as well as their frequency. Using that
information, it is possible to step back in time and infer the conditions that
led to the final configuration of the merger remnant. Previous papers in the
\sauron\ series, concentrated on E and S0 galaxies, have shown that KDCs are
common in early-type galaxies (e.g. de Zeeuw et al. 2002; Emsellem et al. 2004).
In general these are old KDCs that are thought to have formed early in the
evolution of the host galaxies (Kuntschner et al. 2006). By zooming in onto the
nuclear regions, however, one discovers that there is also a large fraction of
young decoupled components in the centres (see Mcdermid et al., these
proceedings). 

\begin{figure}[!t]
\begin{center}
\includegraphics[width=0.49\linewidth]{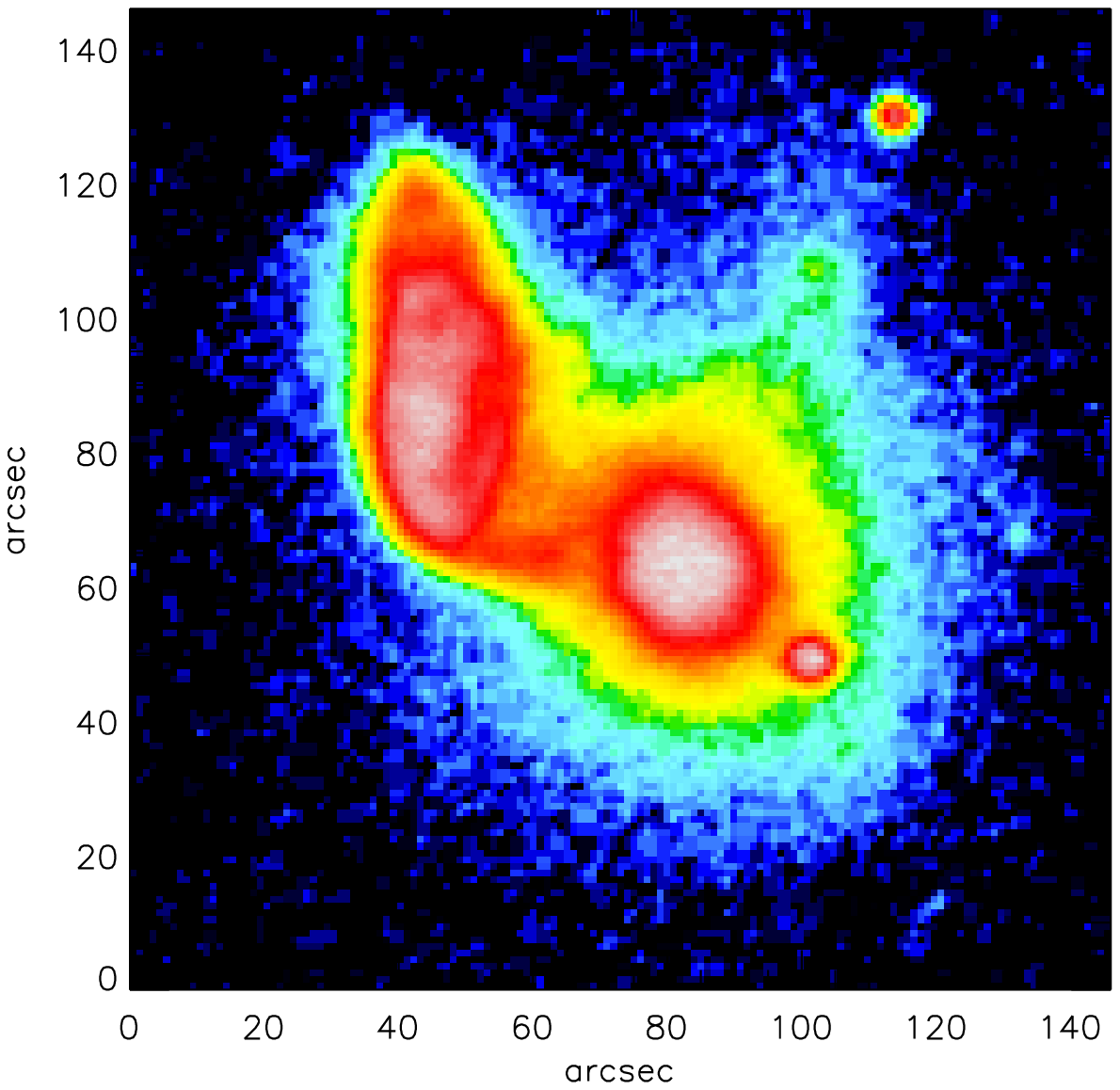}
\includegraphics[width=0.49\linewidth]{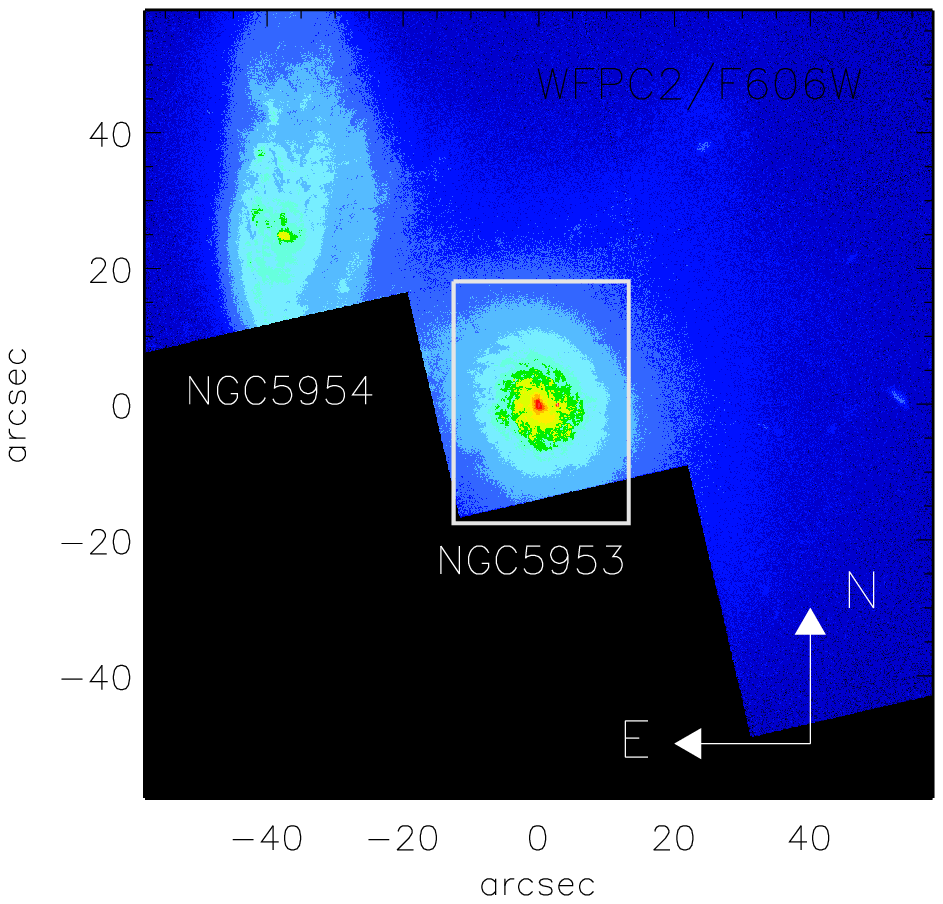}
\caption{The interacting pair formed by NGC\,5953 and NGC\,5954. Left: $B$-band
DSS image (logarithmically scaled) showing the bridge of material between the
 two galaxies. Right: High spatial resolution HST/WFPC2-F606W image. Overlaid on
NGC\,5953 is the \sauron\ field-of-view covering the inner 4$\times$3 kpc of the
galaxy.}
\end{center}
\end{figure}

Among the 24 Sa spiral bulges in the \sauron\ survey, we were able to detect
kinematically decoupled components in up to 50\% of the galaxies, most of them
aligned with the galaxies photometric major axis, and likely the end result of
secular evolutionary processes (e.g. Wozniak et al. 2003). Only two cases
display strongly misaligned or counter-rotating KDCs (i.e. NGC\,4698 and
NGC\,5953, see Falc\'on-Barroso et al. 2006). NGC\,4698 is a well studied
galaxy, where the KDC may be the result of an intermediate-size merger (Bertola
et al. 1999). The case of NGC\,5953 is particularly interesting in the survey
because it is the only example where we are witnessing an interaction that is
currently taking place (see Figure~1), rather than observe the end product of a
past merger.

\begin{figure}[!t]
\begin{center}
\includegraphics[width=0.99\linewidth]{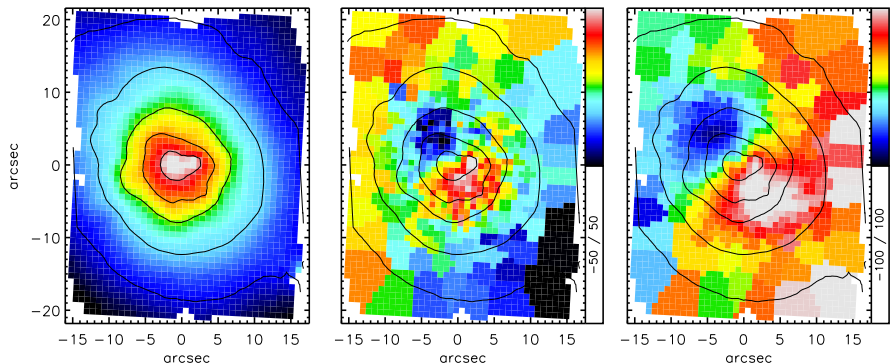}
\caption{Kinematics of NGC\,5953. From left to right: i) \sauron\ reconstructed
intensity, ii) stellar velocity, iii) \hbeta\ ionised-gas velocity. Levels for
the kinematics are indicated in a box on the right-hand side of each map, and
are expressed in \kms.}
\end{center}
\end{figure}

%------------------------------------------------------------------
\section{The interacting galaxy NGC\,5953}
The system formed by the interacting pair NGC\,5953/NGC\,5954 has been
extensively studied in the past. Most of the work has focused on the active
nature of NGC\,5953's nucleus (Jenkins 1984; Reshetnikov 1993; Gonzalez Delgado
\& Perez 1996). Rampazzo et al. (1995) combined broad-band imaging and
long-slit spectroscopy to produce a detailed photometric and kinematic
analysis of the interaction between the two galaxies. More recent results make
use of a Fabry-P\'erot spectrograph to map the two-dimensional gas kinematics
using the \nii\ emission line (Hern\'andez et al. 2003; see also Fuentes-Carrera
et al. in these proceedings).\looseness-2

In Figure~2, we present the reconstructed intensity map and the stellar and gas
kinematics of the central regions of NGC\,5953. The \sauron\ observations
reveal the presence of a kinematically decoupled component in the central
regions, where the stars are counter-rotating with respect to the bulk of the
galaxy. The ionised-gas velocity map of the central component is consistent
with that of the stars in the inner parts, and it is in good agreement with that
of Hern\'andez et al. (2003). While the kinematic major axis of the stellar
component (both the inner and outer parts) appears to be aligned with respect to
the photometric major axis of the galaxy, the kinematics of the ionised-gas in
the outer parts shows signs of non-axisymmetry.

In addition to the stellar and the gas kinematics, we have also determined the
\oiii/\hbeta\ ratio over the \sauron\ field-of-view. This ratio is a good
indicator of young metal-rich stellar populations where star formation is very
intense (Kauffmann et al. 2003). In Figure~3 (left panel) the ratio map exhibits
very low values at the location of the kinematically decoupled component and
high values in the very centre of the galaxy. These results are similar to those
of Yoshida et al. (1993) in the inner regions (i.e. \oiii/\hbeta=0.14-0.41) and
are consistent with a star-forming region surrounding an active galactic
nucleus.

\begin{figure}[!t]
\begin{center}
\includegraphics[width=0.99\linewidth]{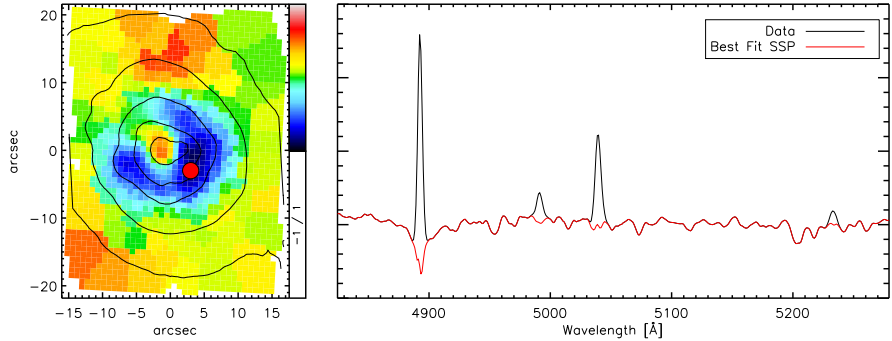}
\caption{A young KDC in NGC\,5953. From left to right: i) \sauron\ \oiii/\hbeta\
ratio map (in logarithmic scale), ii) \sauron\ spectrum (black solid line) at
the location marked in the left panel (red dot), with the best single stellar
population model (t=3~Gyr) overlaid (red solid line).}
\end{center}
\end{figure}

%------------------------------------------------------------------
\section{A recently formed KDC in NGC\,5953}
Despite the fact that the gas and stars in the inner parts of the galaxy
appear to be kinematically decoupled from the main body and despite the
extremely low \oiii/\hbeta\ ratio in the same location, there is no direct proof
that both the star-forming gas and stellar decoupled component are related,
although this is likely. To establish this link, we need to estimate the age of
the stars in that region and determine whether it is indeed consistent with a
young stellar population.\looseness-2

Here we follow the simple procedure outlined in Vazdekis (1999) to measure the
age of the underlying stellar population. Briefly, the idea is to find the
single stellar population (SSP) model from a library of model templates that
best matches the overall spectrum in the kinematically decoupled region.
Figure~3 (right panel) shows the result of this experiment. The best-fit SSP
model has an age of $\sim$3~Gyr. By making the simple assumption that the
galaxy is made by two populations (i.e. young and old), then the
luminosity-weighted age of 3 Gyr measured here sets an upper limit to the
age of the young population. The true age of this population is however
difficult to assess given that different fractions of young stars on top of an
old population can lead to the same luminosity-weighted age. Nevertheless,
given the quality of the fit, it seems certain that young stars are present at
the location of the KDC.\looseness-2

The combination of these different observations suggests that at least part of
the gas in NGC\,5953 was acquired from the nearby interacting galaxy NGC\,5954,
settled to the disk plane with a direction of rotation opposite to
that of the stars in the main disk, and itself started to form stars. The
kinematic and population information set strong constraints on the formation
timescale of the KDC we observe in this galaxy, and suggest that we are indeed
witnessing the early stage of its formation.

%------------------------------------------------------------------

%%%%%%%%%%%%%%%%%%%%%%%%%%%%%%%%%%%%%%%%%%%%%%%%%%%%%%%%%%%%%%%%%%%%%%
\printindex
\end{document}